\documentclass[aps,notitlepage,oneside,superscriptaddress]{revtex4-1}
\usepackage{latexsym}
\usepackage{amssymb}
\usepackage{amsmath}

\begin{document}
\title{Double layer in quadratic gravity and least action principle} 

\author{V. A. Berezin}\thanks{berezin@inr.ac.ru}
\affiliation{Institute for Nuclear Research of the Russian Academy of Sciences, prospekt 60-letiya Oktyabrya 7a, Moscow 117312, Russia}
\author{V. I. Dokuchaev}\thanks{dokuchaev@inr.ac.ru}
\affiliation{Institute for Nuclear Research of the Russian Academy of Sciences, prospekt 60-letiya Oktyabrya 7a, Moscow 117312, Russia}
\author{Yu. N. Eroshenko}\thanks{eroshenko@inr.ac.ru}
\affiliation{Institute for Nuclear Research of the Russian Academy of Sciences, prospekt 60-letiya Oktyabrya 7a, Moscow 117312, Russia}
\author{A. L. Smirnov}\thanks{smirnov@ms2.inr.ac.ru}
\affiliation{Institute for Nuclear Research of the Russian Academy of Sciences, prospekt 60-letiya Oktyabrya 7a, Moscow 117312, Russia}

\date{\today}

\begin{abstract}
The Israel equations for thin shells in General Relativity are derived directly from the least action principle. The method is elaborated for obtaining the equations for double layers in quadratic gravity from the least action principle.
\end{abstract}
\keywords{General Relativity \and quadratic gravity \and double layer \and least action principle}
\maketitle \tableofcontents

\section{Introduction}

Any relativistic gravitation theory should be described by nonlinear equations, since the gravitational field has energy and consequently is a source itself. The solution of these equations is extremely challenging even in the case of vacuum. Therefore,
singular distributions of the energy-momentum tensor of material (nongravitational) sources are of special interest. We mean primarily the distributions described by the Dirac $\delta$-function, i.e., thin shells.

More specifically, we consider the action integral
in the form of a sum of gravitational action $S_{\rm grav}$, and
the actions for the matter fields, $S_{\rm m}$,
\begin{equation}
	S_{\rm tot}=S_{\rm grav}+S_{\rm m}.
\end{equation}
According to the definition, the variation of the matter
action as a metric tensor $g_{\mu\nu}$ ($ds^2=g_{\mu\nu}dx^\mu dx^\nu$) gives us the energy-momentum tensor $T_{\mu\nu}$:
\begin{equation}
	\delta S_{\rm m}\stackrel{\mathrm{def}}{=} \frac{1}{2}\int\!T_{\mu\nu}\delta g^{\mu\nu}\sqrt{-g}\,d^4x=\frac{1}{2}\int\! T^{\mu\nu}\delta g_{\mu\nu}\sqrt{-g}\,d^4x,
\end{equation}
where $g$ is the determinant of tensor $g_{\mu\nu}$.

It is assumed that there is a singular surface $\Sigma_0$ inside the integration domain, on which the energy-momentum tensor is localized. We are interested only in this singular part. The three-dimensional hypersurface  $\Sigma_0$ divides the four-dimensional spatiotemporal integration domain into two parts, conventionally internal (``$-$'') and external (``$+$''). The transformation of coordinates in each of these domains can provide the way to attain the metric tensor continuity on $\Sigma_0$ (this is the only thing that connects them). Note that except this continuity, the internal and external parts should be considered absolutely separate and unrelated manifolds. In particular, it is possible to introduce the Gaussian normal coordinate system related with $\Sigma_0$: $x^\mu=(n,x^i)$, $x^i\in\Sigma_0$, 
\begin{equation}
	ds^2=\epsilon\,dn^2+\gamma_{ij}dx^idx^j,
\end{equation}
where the coordinate $n$ is directed along the external normal to $\Sigma_0$, and $\epsilon=\pm1$, depending on whether is $\Sigma_0$ a space-like or a time-like hypersurface (we use signature $(+---)$), and the equation for $\Sigma_0$ is just $n=0$. The enclosure $\Sigma_0$ into the four-dimensional volume is described by the extrinsic curvature tensor $K_{ij}$:
\begin{equation}
K_{ij}\stackrel{\mathrm{def}}{=} 
-\frac{1}{2}\frac{\partial\gamma_{ij}}{\partial n}\Big|_{\Sigma_0}.
\end{equation}
In these coordinates $T_{\mu\nu}=S_{\mu\nu}\delta(n)+\dots$ and 
\begin{equation}
\delta S_{\rm m}=\frac{1}{2}\int\limits_{\Sigma_0}\! 
(S_{nn}\delta g^{nn}+2S_{ni}\delta g^{ni}
+S_{ij}\delta g^{ij})\sqrt{|\gamma|}\,d^3x.
\end{equation}
Tensor $S_{\mu\nu}$ is the energy-momentum surface tensor on
the mass shell. Note that, although $g_{nn}=g^{nn}=\epsilon$, $g_{ni}=g^{ni}=0$, their variations are not necessarily equal to zero on $\Sigma_0$.

We shall work with the Riemann geometry, i.e., the connectivity coefficients $\Gamma^\lambda_{\mu\nu}$ can be written as
\begin{equation}
\Gamma_{\mu\nu}^\lambda=\frac{1}{2} g^{\lambda\sigma}(g_{\sigma\mu,\nu}+g_{\sigma\nu,\mu}-
g_{\mu\nu,\sigma}),
\end{equation}
where comma denotes the partial derivative. The Riemann curvature tensor is
\begin{equation}
	R^\mu_{\phantom{1}\nu\lambda\sigma}=\frac{\partial\Gamma^\mu_{\nu\sigma}}{\partial x^\lambda}
	-\frac{\partial\Gamma^\mu_{\nu\lambda}}{\partial x^\sigma}
	+\Gamma^\mu_{\varkappa\lambda}\Gamma^\varkappa_{\nu\sigma}
	-\Gamma^\mu_{\varkappa\sigma}\Gamma^\varkappa_{\nu\lambda},
\end{equation}
and the Ricci tensor $R_{\nu\sigma}$ and the curvature scalar $R$ are its convolutions: and $R_{\nu\sigma}=R^\mu_{\phantom{1}\nu\mu\sigma}$, $R=g^{\mu\nu}R^\mu_\mu$. In the Gaussian normal coordinates near the hypersurface $\Sigma_0$ these formulas can be written as follows (vertical bar
denotes the covariant derivative in the three-dimensional
metric $\gamma_{ij}$):
\begin{equation}
	\Gamma_{ij}^n=\epsilon K_{ij}, \quad \Gamma_{ni}^l=- K^l_i, \quad
	\Gamma_{ij}^l=^{(3)\!\!}\Gamma_{ij}^l,
\end{equation}
\begin{equation}
	R^n_{\phantom{1}inj}=\epsilon (K_{ij|n}+K_{lj}K_i^l),
\end{equation}
\begin{equation}
	R^n_{\phantom{1}ikj}=\epsilon (K_{ij|k}+K_{ik|j}),
\end{equation}
\begin{equation}
	R^l_{\phantom{1}ikj}=-\epsilon (K_{ij}K_k^l-K_j^l K_{ik})
	+^{(3)\!\!}R^l_{\phantom{1}ikj},
\end{equation}
\begin{equation}
	R_{nn}=K_{ij|n}+K_{lj}K_i^l, \quad R_{ni}=-K_{i|j}^j+K_{|i},
\end{equation}
\begin{equation}
	R_{ij}=\epsilon(K_{ij|n}+K_{lj}K_i^l)+^{(3)\!\!}R_{ij},
\end{equation}
\begin{equation}
	R=2\epsilon(K_{,n}+K_{lj}K^i_lK^l_i)+^{(3)\!\!}R,
\end{equation}
\begin{equation}
	[R_{ninj}]=[K_{ij,n}],  \quad [R_{nn}]=\gamma^{ij}[K_{ij,n}],
\end{equation}
\begin{equation}
	[R_{ij}]=\epsilon[K_{ij,n}],  \quad [R]=2\epsilon\gamma^{ij}[K_{ij,n}].
\end{equation}
The components of the metric tensor $g_{\mu\nu}$ or the inverse tensor $g^{\mu\nu}$ are the only dynamic variables in the gravitational action.

We assume that the singular surface  $\Sigma_0$ is fixed. The variation of the total action should be zero independently in the external volume and in the internal volume (we can always choose arbitrarily an ``auxiliary'' integration volume that does not affect the singular hypersurface), which gives us the field equations. The equality to zero of the variation of action on  $\Sigma_0$ leads to the equations for matching the solutions in the (``$+$'') and (``$-$'') domains.

In General Relativity, as we know, the gravitational field equations are second order equations with
respect to the metric tensor derivatives, despite the presence of these second derivatives in the Hilbert Lagrangian proportional to the curvature $R$ scalar. Therefore, the appearance of a thin shell in the form of a $\delta$-function in the energy-momentum tensor (on the right-hand side of the Einstein equations) should be necessarily offset by a similar $\delta$-function in the second derivatives of the metric, and consequently in the curvature scalar. Hence, we obtain the following logic chain: the $\delta$-function in the second derivatives of the metric jump in the first derivatives continuity ofthe metric tensor. The matching equations determine the jump of the external curvature when passing across the singular hypersurface $\Sigma_0$. These equations were first obtained by W. Israel \cite{israel}. The Israel equations described not only thin shells, but also shock waves when a jump of the energy-momentum tensor of matter occurs. In this case, the external curvature tensor is continuous on $\Sigma_0$, while its normal derivatives (and the metric tensor second derivatives, respectively) and the curvature scalar undergo a discontinuity $R$. This means that the shock wave of matter is accompanied by a gravitational shock wave. In the next section, we
obtain the Israel equations from the least action principle, instead of direct integration of the field equations.

In quadratic gravitation, the situation with a singular surface is not so simple. The Lagrangian of quadratic gravitation is a sum of squared Riemann curvature tensors, Ricci tensor, curvature scalar, curvature scalar linear in curvature, and cosmological constant. Therefore, the field equations contain the metric tensor derivatives up to the fourth order inclusive. Unlike General Relativity, the extrinsic curvature tensor of the singular hypersurface $\Sigma_0$ is bound to be continuous, otherwise we would have obtained the $\delta$-function in the curvature and the $\delta^2$-function in the action integral, which is strongly forbidden in the standard theory of generalized functions (to which we adhere). Therefore, the curvature can have not more than a bound on the $\Sigma_0$. In that case, the logic chain is the following: the second derivatives of the metric tensor undergo a jump the third derivatives have a singularity in the form of $\delta$-functions the fourth derivatives have a singularity in the form of a $\delta'$-function. This is traditionally called a double layer. The general equations of the double layer in quadratic gravitation were obtained by J.M. Senovilla \cite{senovilla}. We have investigated in detail the case of spherical symmetry in Weyl$+$Einstein gravitation \cite{berezin18} and have found some interesting features hidden in the general formalism. The purpose of this paper is to derive the double layer equations from the least action principle and to clarify the nature of these features. 

Hence, a bound in the curvature causes the appearance of a double layer in quadratic gravitation. In the General Relativity, this jump means the emergence of a shock gravitational wave accompanied by a shock wave of matter fields. Now, in the gravitational field equations, a $\delta'$-function, a $\delta$-function, and the jump in derivatives of the metric tensor appear, while in the energy-momentum tensor, a $\delta$-function ($=$ thin shell) and the jump ($=$ shock wave) emerge. This means that in quadratic gravitation, a shock gravitational wave can accompany a thin shell, though it can exist without it. (Note that a shock wave in the matter fields may not cause a shock gravitational wave).

Notations: comma ``,'' preceding the index denotes the ordinary partial derivative, semicolon ``;'' denotes the covariant derivative in the four-dimensional metric $g_{\mu\nu}$, and vertical bar ``|'' denotes the covariant derivative in the three-dimensional metric $\gamma_{ij}$. An expression in square brackets denotes the jump across the hypersurface  $\Sigma_0$, i.e., $[\ldots]\equiv (+)-(-)$.

\section{Deriving the Israel equations from the least action principle}

We start from the Hilbert action in the General
Relativity:
\begin{equation}
	S_{\rm H} = -\frac{1}{16\pi G}\int\! R\sqrt{-g}\,d^4x.
\end{equation}
The integration is carried out over the four-dimensional volume confined by the hypersurface . Here,
we omit an auxiliary surface term that is needed, as we know, to observe the least action principle at fixed-edge variations (on  $\Sigma$). What is important for us here is that the extrinsic curvature tensor variation $\delta K_{ij}$ is completely arbitrary on $\Sigma$, whereas $\delta K_{ij}$ are equal to zero by definition.

Let a certain given hypersurface  $\Sigma_0$ exists inside the variation volume, on which a part of the matter energy-momentum tensor proportional to the Dirac $\delta$-function is localized. Subject to the Einstein equations, a similar term appears in the scalar curvature $R$ included in the action integral. Then we have two options. We can first vary the action, and then integrate the $\delta$-function. We applied this approach in \cite{berezin19}. Here, we take another way: first, we integrate the function, and then we vary the action. Thus, we have
\begin{equation}
	S_{\rm H} = \frac{1}{16\pi G}\left\{+\int\limits_{\Sigma_0} 2[K]\sqrt{|\gamma|}\,d^3x +\int\limits_{(\pm)} R\sqrt{-g}\,d^4x \right\}.
\end{equation}
The variation of this action is
\begin{eqnarray}
\delta S_{\rm H} &=& \frac{1}{16\pi G}\Biggl\{\epsilon\!\int\limits_{\Sigma_0} \left(2[\delta K]
-[K]\gamma_{ij}\delta\gamma^{ij} \right)\sqrt{|\gamma|}\,d^3x \Biggr.\nonumber \\
&&+\int\limits_{(\pm)}\!\Biggl.
\left(\gamma^{\mu\nu}\delta R_{\mu\nu}
+\Bigl(R_{\mu\nu}-\frac{1}{2}g_{\mu\nu}\Bigr)\delta\gamma^{\mu\nu}\right)
\sqrt{-g}\,d^4x\Biggr\}.
\end{eqnarray}
Here we give a brief description of the procedure adopted. We aim to obtain the equations on the singular hypersurface $\Sigma_0$, which would relate the solutions in the external $(+)$ and internal $(-)$ domains. Therefore, we assume the ``initial'' conditions for some solutions with similar $\gamma_{ij}$, but different (arbitrary) $K_{ij}$ (this is possible due to an auxiliary surface term) set on the boundary hypersurface, which induce $\delta\gamma_{ij}$ and $\delta K_{ij}$on the singular hypersurface $\delta K_{ij}$. Then
\begin{equation}
	\delta K_{ij} = A_{ijlp}\delta\gamma^{lp},
\end{equation}
since these variations depend on the chosen solutions on the boundary hypersurface, but the tensor $A_{ijlp}$ is completely arbitrary (without taking into consideration the symmetry inside each pair of indices).

Since the Einstein equations hold outside $\Sigma_0$, only the terms $\delta R_{\mu\nu}$ proportional to remain from the volume integrals, and, according to the remarkable formula $\delta R^\mu_{\phantom{1}\nu\lambda\sigma} =(\delta T^\mu_{\nu\sigma})_{;\lambda}
-(\delta\Gamma^\mu_{\nu\lambda})_{;\sigma}$ (the semicolon means a covariant derivative with respect to the four-dimensional metric ), they are easily converted into a total derivative and, subject to the Stokes theorem, take the form
\begin{eqnarray}
	&-&\int\limits_{\Sigma_0}g^{\mu\nu} [\delta\Gamma^\lambda_{\mu\nu}]\sqrt{-g}\,dS_\lambda 
	+ \int\limits_{\Sigma_0}g^{\mu\nu} [\delta\Gamma^\lambda_{\mu\lambda}]\sqrt{-g}\,dS_\nu 
	\nonumber \\
	&=&\int\limits_{\Sigma_0}\left(-g^{\mu\nu}[\delta\Gamma^n_{\mu\nu}]
	+\epsilon[\delta\Gamma^\lambda_{n\lambda}] \right) \sqrt{|\gamma|}\,d^3x
	\nonumber \\
	&=&-\epsilon\int\limits_{\Sigma_0}\left(\gamma^{ij}[\delta K_{ij}]
	+[\delta K] \right)\sqrt{|\gamma|}\,d^3x.
\end{eqnarray}
While the variation of full action $\delta S_{\rm H}+\delta S_{\rm m}$ is
\begin{equation}
	\frac{\epsilon}{16\pi G}\int\limits_{\Sigma_0}\left([K_{ij}]
	-\gamma_{ij}[K]\delta\gamma^{ij}\right)\sqrt{|\gamma|}\,d^3x
	+\frac{1}{2}\int\limits_{\Sigma_0}S_{\mu\nu}\delta g^{\mu\nu}\sqrt{|\gamma|}\,d^3x=0.
\end{equation}
Note that the arbitrary variation $\delta K_{ij}$ has completely disappeared in the final expression. In addition, the summation in the second integral is carried out over all four indices, since  $\delta g^{nn}$ and $\delta g^{ni}$ must not be equal to zero. The absence of these terms in the first integral means that $S_{nn}=S_{ni}=0$.

Here we reproduce the Israel equations:
\begin{equation}
	\epsilon([K_{ij}]-\gamma_{ij}[K])=8\pi G S_{ij}, \quad S_{nn}=0,
	\quad S_{ni}=0.
\end{equation}

\section{Double layer in quadratic gravitation}

Finally, let us address the primal problem, deriving the equations for a double layer in quadratic gravitation, which define the matching of solutions in the external and internal regions of the four-dimensional volume resolved by the three-dimensional singular hypersurface $\Sigma_0$. All the time, we underline the differences from General Relativity.

The gravitational part of the action $S_2$ is
\begin{equation}
	S_2=\int\!{\cal L}_2\sqrt{-g}\,d^4x,
\end{equation}
where
\begin{equation}
	{\cal L}_2=\alpha_1R_{\mu\nu\lambda\sigma}R^{\mu\nu\lambda\sigma}
	+\alpha_2R_{\mu\nu}R^{\mu\nu}+\alpha_3R^2+\alpha_4\Lambda.
\end{equation}
By contrast with General Relativity, where the $\delta$-function in the energy-momentum tensor caused, necessarily, the occurrence of the $\delta$-function in the curvature scalar $R$ and, consequently, in the Lagrangian, now, as already mentioned in the Introduction, the curvature can only display a jump on $\Sigma_0$. And this jump results in the appearance of a gravitational double layer. The variation of the action $S_2$ is
\begin{eqnarray}
\delta{\cal L}_2&=&\int\left\{2\alpha_1R_{\mu}^{\phantom{1}\nu\lambda\sigma}\delta R^{\mu}_{\phantom{1}\nu\lambda\sigma}
+2\alpha_2R^{\mu\nu}\delta R_{\mu\nu}
+2\alpha_3Rg^{\mu\nu}\delta R_{\mu\nu}
+\alpha_4g^{\mu\nu}\delta R_{\mu\nu}\right. \nonumber \\
&+&\bigl.(\ldots)\,\delta g^{\mu\nu}\bigr\}\sqrt{-g}\,d^4x.
\end{eqnarray}
We have not written a long series of terms in round brackets, because their sum is zero in the final expression following from the field equations in the external and internal volumes, while the $\delta$-function has not appeared yet.

Further,
\begin{eqnarray}
	\delta{\cal L}_2&\rightarrow&\int\left\{2\alpha_1R_{\mu}^{\phantom{1}\nu\lambda\sigma}
	\bigl( (\delta\Gamma^{\mu}_{\nu\sigma})_{;\lambda}
	-(\delta\Gamma^{\mu}_{\nu\lambda})_{;\sigma} \bigr) \right.\nonumber \\
	&+&\left.(2\alpha_2R^{\mu\nu}+2\alpha_3Rg^{\mu\nu}
	+\alpha_4g^{\mu\nu})\left((\delta\Gamma^{\lambda}_{\mu\nu})_{;\lambda}
	-(\delta\Gamma^{\lambda}_{\mu\lambda})_{;\nu}\right)\right\}\sqrt{-g}\,d^4x \nonumber \\
	&=&\int\left\{(2\alpha_1R_{\mu}^{\phantom{1}\nu\lambda\sigma}
	\delta\Gamma^{\mu}_{\nu\sigma})_{;\lambda}
	-(2\alpha_1R_{\mu}^{\phantom{1}\nu\lambda\sigma}
	\delta\Gamma^{\mu}_{\nu\lambda})_{;\sigma}\right. \nonumber \\
	&+&\left. \left((2\alpha_2R^{\mu\nu}+2\alpha_3Rg^{\mu\nu}
	+\alpha_4g^{\mu\nu})\delta\Gamma^{\lambda}_{\mu\nu}\right)_{;\lambda}
	-\left((2\alpha_2R^{\mu\nu}+2\alpha_3Rg^{\mu\nu}
	+\alpha_4g^{\mu\nu})\delta\Gamma^{\lambda}_{\mu\lambda}\right)_{;\nu}
	\right.  \nonumber \\
	&-&\left.2\alpha_1\left(R^{\phantom{1}\nu\lambda\sigma}_{\mu\phantom{111};\lambda}   \delta\Gamma^{\mu}_{\nu\sigma}
	-R^{\phantom{1}\nu\lambda\sigma}_{\mu\phantom{111};\sigma}
	\delta\Gamma^{\mu}_{\nu\lambda}  \right)
	-(2\alpha_2R^{\mu\nu}+2\alpha_3Rg^{\mu\nu}
	+\alpha_4g^{\mu\nu})_{;\lambda}\delta\Gamma^{\lambda}_{\mu\nu} \right. \nonumber \\
	&+&\left. (2\alpha_2R^{\mu\nu}+2\alpha_3Rg^{\mu\nu}
	+\alpha_4g^{\mu\nu})_{;\nu}\delta\Gamma^{\lambda}_{\mu\lambda}
	\right\} \sqrt{-g}\,d^4x  
\end{eqnarray}
It is necessary to note a very important difference from General Relativity here. Since $\delta$-functions are absent in the Lagrangian $\delta{\cal L}_2$, there are no functions in the variations, and the ``explicit'' appearance of these $\delta$-functions in the covariant derivatives of the Riemann tensor and its convolutions is totally recouped by the functions ``hidden'' in the total derivatives. Hence, we can safely consider the volume integral as a sum of integrals, separately over the ``$+$'' and ``$-$'' domains. The same is true for the $\delta'$-function. It is especially instructive to consider a surface integral arising from total derivatives with the participation of variations of the connectivity coefficients $\delta\Gamma$; we denote this integral $\delta\Sigma_1$:
\begin{eqnarray}
\delta{\cal L}_2&\rightarrow&-\int\limits_{\Sigma_0}\left\{4\alpha_1[R_{\mu}^{\phantom{1}\nu n\sigma}]\delta\Gamma^{\mu}_{\nu\sigma}
+2(\alpha_2[R^{\mu\nu}]+\alpha_3[R]g^{\mu\nu})\delta\Gamma^{n}_{\mu\nu}
\right.\nonumber \\
&-&\left.2(\alpha_2[R^{\mu n}]+\alpha_3[R]g^{\mu n})\delta\Gamma^{\lambda}_{\mu\lambda} \right\}\sqrt{|\gamma|}\,d^3x. 
\end{eqnarray}
Here we have used the symmetries of the curvature tensor and the equality to zero of a covariant derivative of the metric tensor and its continuity on $\Sigma_0$. Note that the coefficients$\alpha_1$ and $\alpha_2$ disappear already at this stage. It means that there is no smooth transition from quadratic gravitation to Einstein gravitation. The general sign in front of the integral is stipulated by the direction of the external normal in respect of the
hypersurface $\Sigma_0$ (from ``$-$'' domain to ``$+$'' domain) and the jump definition ($[\ldots]=(+)-(-)$). Substituting the values of jumps for the Riemann tensor and its convolutions determined by the jump of the normal derivative of the external curvature tensor $[K_{ij,n}]$ we obtain:
\begin{eqnarray}
	\delta{\cal L}_2&\rightarrow& -\int\limits_{\Sigma_0}\Bigl\{\left((4\alpha_1+\alpha_2)\gamma^{il}\gamma^{jp}+(\alpha_2+4\alpha_3)\gamma^{ij}\gamma^{lp}\right)[K_{ij:n}]\delta K_{lp}
	\Bigr.  \\
	&+&\left.\left(2\alpha_1K^j_l[K_{pj;n}]+(\alpha_2
	+2\alpha_3)\gamma^{ij}K_{lp}[K_{ij;n}]\right)\delta\gamma^{lp}\right\}
	\sqrt{|\gamma|}\,d^3x.
\end{eqnarray}
The nonzero jump of the normal derivative of the extrinsic curvature tensor $K_{ij,n}$ is necessary for the existence of a gravitational double layer. We see that by contrast with General Relativity, a generally irremovable variation of the external curvature tensor $\delta K_{lp}$ with completely arbitrary coefficient $[K_{ij,n}]$ emerges in quadratic gravitation, except in the case when $\alpha_2=-4\alpha_1$ and $\alpha_3=\alpha_1$. However, this is in line with the Gauss-Bonnet term that does not influence the field equations in the four-dimensional metric, so that there is not a trace left of the double layer.

As for the general case, the external curvature tensor variations $\delta K_{ij}$ are not recouped at all, by contrast with General Relativity. Therefore, as already mentioned above,
\begin{equation}
	\delta K_{ij} = B_{ij}^{\mu\nu}\delta\gamma_{\mu\nu},
\end{equation}
where $B_{ij}^{\mu\nu}$ is the arbitrary four-rank tensor symmetric with respect to each pair of indices. It should be determined when solving the equations of matching on the singular hypersurface . The appearance of such ``arbitrariness'' was first mentioned by the authors in\cite{berezin19} in an example of Weyl$+$Einstein spherically symmetric gravitation. It appeared there as a result of mathematical operations with the $\delta$-function derivative, $\delta'$. Here, we proceed without it.

The remaining volume integral is
\begin{eqnarray}
	\delta{\cal L}_2&\rightarrow&\int\bigl\{4\alpha_1R^{\phantom{1}\nu\lambda\sigma}_{\mu\phantom{111};\sigma}   \delta\Gamma^{\mu}_{\nu\lambda}
	-2(\alpha_2R^{\mu\nu}_{;\lambda}
	+\alpha_3R_{;\lambda}g^{\mu\nu})\delta\Gamma^{\lambda}_{\mu\nu}
	\bigl. \nonumber \\
	&+&\left. 2(\alpha_2R^{\mu\nu}_{\phantom{11};\nu}+\alpha_3R_{;\nu}g^{\mu\nu}
	)\delta\Gamma^{\lambda}_{\mu\lambda}
	\right\} \sqrt{-g}\,d^4x.  
\end{eqnarray}
There is another remarkable formula for the connectivity coefficient variations:
\begin{equation}
\delta\Gamma^\lambda_{\mu\nu}=
\frac{1}{2}g^{\lambda\sigma}\bigl((\delta g_{\sigma\mu})_{;\nu}
+(\delta g_{\sigma\nu})_{;\mu} -(\delta g_{\mu\nu})_{;\sigma}\bigr). 
\end{equation}
Further, we extract the total derivatives and applying the Stokes theorem obtain another contribution to the surface integral, which we denote $\delta\Sigma_2$:
\begin{eqnarray}
\delta\Sigma_2&=&\int\limits_{\Sigma_0}\left\{-4\alpha_1[R^{\alpha n\lambda\sigma}_{\phantom{1111};\sigma}]\delta g_{\alpha\sigma}
+\alpha_2\left(2[R^{\mu n;\alpha}]\delta g_{\mu\alpha}-[R^{\mu\nu;n}]\delta g_{\mu\nu}-\frac{1}{2}[R^{;n}]g^{\lambda\sigma}\delta g_{\lambda\sigma}\right)
\right.\nonumber \\
&+&\biggl. 2\alpha_3([R^{;i}]\delta g_{ni}-[R^{;n}]\gamma^{ij}\delta\gamma_{ij} )\biggr\}\sqrt{|\gamma|}\,d^3x.  
\end{eqnarray}
Further evaluations are complete routine. Let us describe qualitatively the basic features of the result. Firstly, this is the appearance of arbitrary functions in the equations of matching, which are defined individually for each pair of the matched solutions. Secondly, which was mentioned by Senovilla \cite{senovilla}, generally speaking, the components $S^{nn}$ and $S^{ni}$ of the surface energy-momentum tensor of matter on the singular shell of $\Sigma_0$ are nonzero (by contrast with General Relativity). We have obtained a stronger result:  $S^{nn}$ and $S^{ni}$ do not disappear even in the case when the quadratic Lagrangian is the Gauss-Bonnet combination, which is a total derivative not influencing the field equations in the volume. The situation becomes quite
different on a singular hypersurface. And this issue still remains to be dealt with.

\section*{Acknowledgments}

We are grateful to E. O. Babichev for stimulating discussions. The paper was supported by the Russian Foundation for Basic Research, project no. 18-52-15001-NCNIa.

\end{document}